\documentclass[sn-mathphys,Numbered]{sn-jnl}
\usepackage{graphicx}
\usepackage{color}
\usepackage{amssymb}
\usepackage{amsfonts}
\usepackage{mathtools}
\usepackage{tikz} 
\usepackage{pgfplots}

\usepackage{graphicx}%
\usepackage{multirow}%
\usepackage{amsmath,amssymb,amsfonts}%
\usepackage{amsthm}%
\usepackage{mathrsfs}%
\usepackage[title]{appendix}%
\usepackage{xcolor}%
\usepackage{manyfoot}%

\newcommand{\Q}{\mathcal Q}

\newcommand{\R}{\mathcal R}

\newcommand{\A}{\mathcal A}
\newcommand{\C}{\mathcal C}

\newtheorem{lemma}{Lemma}
\newtheorem{corollary}{Corollary}
\newtheorem{proposition}{Proposition}

\newtheorem{example}{Example}

\newcommand{\blue}{\textcolor{black}}

\begin{document}

\title[Interior operators and their relationship to autocatalytic networks]{Interior operators and their relationship to autocatalytic networks}

\author[1]{\fnm{Mike} \sur{Steel}}

\affil[1]{\orgdiv{Biomathematics Research Centre}, \orgname{University of Canterbury},  \city{Christchurch}, \country{New Zealand}}


\abstract{
\blue{The emergence of an autocatalytic network from an available set of elements is a fundamental step in early evolutionary processes, such as the origin of metabolism.} Given the set of elements, the reactions between them (chemical or otherwise), and with various elements catalysing certain reactions, a Reflexively Autocatalytic F-generated (RAF) set is a subset {\em R}$'$ of reactions that is self-generating from a given food set, and with each reaction in {\em R}$'$ being catalysed from within {\em R}$'$.  RAF theory has been applied to various phenomena in theoretical biology, and a key feature of the approach is that it is possible to efficiently identify and classify RAFs within large systems. This is possible because RAFs can be described as  the (nonempty) subsets of the reactions that are the fixed points of an (efficiently computable) interior map that operates on subsets of reactions. Although the main generic results concerning RAFs can be derived using just this property, we show that for systems with at least 12 reactions there are generic results concerning RAFs that cannot be proven using the interior operator property alone. }


\maketitle


\section{Introduction}

Discrete graph-theoretic models have been developed to describe the emergence and structure of self-generating autocatalytic reaction networks within a larger network. This approach was pioneered by Stuart Kauffman's modelling of autocatalytic systems in a simple polymer-based origin-of-life model \cite{kau, kau2}, as well as independent results on the appearance of cycles in random directed graphs \cite{bol, coh} motivated by their relevance to the emergence of living systems.  Kauffman's  notion of a self-generating autocatalytic network was later formalised as a the concept of a Reflexively Autocatalytic and F-generated set (`RAF', defined shortly) \cite{hor}.   The subsequent theory and algorithms concerning RAFs have been applied in a number of areas, ranging from  the origin and structure of primitive metabolism \cite{xav, xav22}, to cognitive modelling in cultural evolution \cite{gab, gab2}, to  ecology \cite{gat18, gat20}, and to economics \cite{gat20}.  The RAF concept is related to (but different from) Robert Rosen's Metabolism-Replacement (M;R) systems in theoretical biology \cite{jar10}.

The task of determining whether or not a large network of `reactions' contains a RAF and if so finding one,  is made tractable (\blue{in polynomial time}) by the property of a certain RAF map defined on the subsets of the full network of reactions.  Here we generalize RAF maps to interior operators and investigate the properties of such operators, as well as the extent to which such operators (on arbitrary finite sets) can be realized as RAF maps. 
In particular, we show that there are generic results concerning RAFs that are not provable from just
the basic properties of the RAF map as an interior operator. \blue{The significance of this result in applications is that certain generic properties of RAFs may require more detailed arguments than those that can be derived using interior operator properties alone.}

We begin by defining interior operators on finite sets, listing some of their basic properties, and describing how they arise naturally from directed graphs. The results are then applied to self-generating autocatalytic networks.

\section{Interior operators and their fixed sets}

In this paper, we will assume that all sets are finite, and given a set $Y$, we write $2^Y$ to denote the power set of $Y$. 
A  function $\psi: 2^Y \rightarrow 2^Y$ is an {\em interior operator} on the subsets of $Y$ if it satisfies the following three properties (nesting, monotonicity, and idempotence) for all subsets $X, X'$ of $Y$:

\begin{itemize}
\item[($I_1$)] $\psi(X) \subseteq X$,
\item[($I_2$)] $X \subseteq X' \Rightarrow \psi(X) \subseteq \psi(X')$, and
\item[($I_3$)]  $\psi(\psi(X)) =\psi(X)$.
\end{itemize}
The  term `interior operator' comes from topology, since the function that assigns to any subspace $S$ of a topological space the interior of $S$ (the union of all the open sets contained in $S$) satisfies the three properties $(I_1)$--$(I_3)$.

Given an interior operator, $\psi: 2^Y \rightarrow 2^Y$ and a subset $X$ of $Y$,  let $$F_\psi(X)  = \{U \subseteq X: \psi(U)=U\}$$  denote collection of subsets of $X$ that are fixed by $\psi$.  We refer to the collection $\{F_\psi(X): X \in 2^Y\}$ as the {\em fixed sets} of $\psi$.  Note that $F_\psi(X) \neq \emptyset$ since $\emptyset \in F_\psi(X)$ for any interior operator $\psi$.

The following lemma summarises some basic and elementary properties of interior operators (a proof is provided in the Appendix). 

\newpage

\begin{lemma}\mbox{}
\label{helps}
 Let, $\psi: 2^Y \rightarrow 2^Y$ be an interior operator, and let $X$ be a subset of $Y$.
\begin{itemize}
\item[(i)] $\psi(X) \in F_\psi(X)$ and $\psi(X)= \bigcup_{U \in F_\psi(X)} U$.
\item[(ii)] $W \subseteq F_\psi(X) \Rightarrow \bigcup W \in F_\psi(X)$.\footnote{\blue{For a collection $W$ of sets we use the shorthand $\bigcup W$ to denote $\cup_{V \in W} V$.}}
\item[(iii)] An arbitrary collection $\C$ of subsets is the collection of fixed sets for some interior operator  if and only if $\emptyset \in \C$  and $\C$ is union-closed.   Moreover, in that case, there is a unique interior operator $\psi_{\C}$ that has $\C$ as its collection of fixed sets, and which is determined by:
\begin{equation}
\label{psiC}
\psi_{\C}(X)= \bigcup{\{U \in \C: U \subseteq X\}},
\end{equation}
for all $X \subseteq Y$.
\end{itemize}
\end{lemma}
Notice that Parts (i) and (ii) of this lemma imply that $\psi(X)$ is the unique maximal fixed set contained within $X$.

\bigskip
      
Next, consider any function $\lambda: 2^Y \rightarrow 2^Y$ that satisfies the properties ($I_1$) and ($I_2$) of an interior operator (but not necessarily $(I_3)$).  Define a function $\psi_\lambda: 2^Y \rightarrow 2^Y$ as follows.
For $X \in 2^Y$,  set 
\begin{equation}\label{lambda}
\psi_\lambda(X) = \bigcap_{i\geq 0} H_i(X),
\end{equation}
where $H_0(X)=X$ and $H_{i+1}(X) = \lambda(H_i(X))$ for all $i\geq 0$.
Notice that since $Y$ is finite, this intersection is finite, and thus, $\psi_\lambda(X) = H_n(X)$ for the first value of $n$ for which $H_n(X)=H_{n+1}(X)$.

 \bigskip
 
\begin{proposition}
\label{firstpro}
If $Y$ is finite, and $\lambda: 2^Y \rightarrow 2^Y$ satisfies the properties ($I_1$) and ($I_2$),  then $\psi_\lambda$ is an interior operator on $2^Y$. Moreover, $\psi_\lambda = \lambda$ if and only if $\lambda$ satisfies ($I_3$).
\end{proposition}

\bigskip

{\em Proof:}
For any $X \in 2^Y$, we have $\psi_\lambda(X) = H_n(X)$ for some value of $n$ (dependent on $X$), and $H_{n+1}(X) = \lambda(H_n(X)) = H_n(X)$.
Thus, $$\psi_\lambda(\psi_\lambda(X))  = H_n(X) \cap \lambda(H_n(X)) \cap \lambda (\lambda(H_n(X)) \cdots$$ Since all of the sets in this intersection equal $H_n(X)$ we obtain
$\psi_\lambda(\psi_\lambda(X))  = H_n(X) =\psi_\lambda(X)$.
For the second claim, if $\lambda =\psi_\lambda$ then since $\psi_\lambda$  satisfies ($I_3$), so does $\lambda$.  Conversely, if $\lambda$ satisfies ($I_3$) then for every $X \in 2^Y$ we have: $$\psi_\lambda(X) = X \cap \lambda(X)  \cap \lambda(\lambda(X)) \cdots = \lambda(X).$$
\hfill$\Box$

\bigskip

\subsection{Interior operators arising from directed graphs}
\label{digraphs}

Let $D=(Y,A)$ be a finite directed graph \blue{with vertex set $Y$}, and for any nonempty subset $X$ of $Y$, let $D|X$ be the induced sub-digraph on $X$ (i.e. $D|X$ has vertex set $X$ and $(u,v)$ is an arc of $D|X$ if and only if  $(u,v) \in A$ and $u,v \in X$).  We let $d_D^+(v)$ denote the in-degree of vertex $v$ in $D$, and for $v \in X$, we  let $d^+_{D|X}(v)$ denote the in-degree of vertex $v$ in $D|X$.  Let $\binom{Y}{k}$ denote the subsets of $Y$ of size $k$, and for 
 $k\geq 1$, let:
$$
\label{eqo}\C_k(D)= \{X \in \binom{Y}{k}: d^+_{D|X}(v) \geq 1 \mbox{ for all } v \in X\}, \mbox{ and }$$
 $$\C(D) = \blue{\{\emptyset\}\cup }\bigcup_{k \geq 1} \C_k(D).$$
 \blue{We say that $\C(D)$ is {\em trivial} if $\C(D)= \{\emptyset\}$}. The following result (particularly Part (iii)) will play an important role in Section~\ref{final}.  The proof is provided in the Appendix.

 \bigskip
 
 \begin{proposition}\mbox{}
\label{propro}
\begin{itemize}
\item[(i)] $\C(D)$ is union-closed;  moreover, $\C(D)$ \blue{is nontrivial} if and only if $D$ contains a directed cycle.
\item[(ii)] If $U, W \in \C(D)$ with $U \subsetneq W$, then either $W\setminus U \in \C(D)$ or there is an element $w \in W\setminus U$ for which $U \cup \{w\} \in \C(D).$
\item[(iii)]  Suppose that $k \geq 3$, $\C_k(D) = \binom{Y}{k}$ and $\C_{j}(D)=\emptyset$ for  all $1\leq j<k$. Then  $$|Y|~\leq~1+(k-1)(k-2).$$
\end{itemize}
\end{proposition}

\bigskip

\noindent{\bf Remarks:} 
\begin{itemize}
\item
In Part (i), the claim that $\C(D)$ \blue{is nontrivial} implies that  $D$ has a directed cycle was noted in \cite{con}.
\item
Proposition~\ref{propro}(iii) fails for $k=1$ or $k=2$; in fact, $Y$ can be arbitrarily large in these cases (e.g., for $k=1$ take the arc set $\{(v,v): v \in Y\}$ and for $k=2$  take the arc set $\{(u,v): u, v \in Y, u \neq v\}$). 

\item
It follows from Proposition~\ref{propro} that not every  interior operator on $2^Y$ can be realised as $\psi_{\C(D)}$ for some  digraph $D$. 
For example, if we let  $Y=\{a,b,c\}$ and take the union-closed set system $\C=\{\blue{\emptyset},  \{a\}, \{a,b,c\}\}$ then  $\C$ cannot equal $\C(D)$ for any digraph $D$ by  Proposition~\ref{propro}(ii).  Alternatively, consider the union-closed set system $\C^+_k = \{X \in 2^Y: |X| \geq k\} \blue{\cup \{\emptyset\}}$, where $k\geq 3$.  This satisfies the two assumptions in Proposition~\ref{propro}(iii) and so for any set $Y$ with  $|Y| > 1+(k-1)(k-2)$ it follows that  $\C^+_k \neq \C(D)$ for any digraph $D$ on vertex set $Y$.  

Moreover, as $Y$ becomes large, the proportion of interior operators on $2^Y$ that can be realised as $\psi_{\C(D)}$ for some $D$ converges to zero as $|Y|$ grows. 
To see this, observe that there are exactly \blue{$2^{n^2}$} digraphs on a vertex set $Y$ of size $n$, and each digraph uniquely determines $\psi_{\C(D)}$ (though many digraphs produce the same interior operator\footnote{For example, by Proposition~\ref{propro}(i), all acyclic digraphs return the trivial interior operator defined by $\psi_{\C(D)}(X) = \emptyset, \forall X \subseteq Y$.}). By contrast, the total number of interior operators on $2^Y$ grows much faster, as the following result shows (a proof is provided in the Appendix). 
\end{itemize}

\begin{proposition}
\label{lem2}
For any set $Y$ of size $n$, there are at least  \blue{$2^{\binom{n}{\lfloor n/2\rfloor}}$} interior operators on $2^Y$.
\end{proposition}

\section{Self-generating autocatalytic networks (RAFs)}

A {\em catalytic reaction system} (CRS) is a quadruple $\Q=(X,R,C,F)$ consisting of a finite nonempty set $X$ of {\em elements} (e.g., molecule types) and a finite set $R$ of {\em reactions}; here a {\em reaction} $r \in R$ refers to an ordered pair $(A,B)$ where $A$ and $B$ are multisets of elements from $X$. In addition,  $C$ is  a \blue{subset  of} $X \times R$ where $(x,r) \in C$ has the interpretation that element $x$ `catalyses' reaction $r$.   We will denote such a CRS by writing $\Q = (X, R, C, F)$.  For each $r \in R$, the subset of $X$ consisting of those elements $x$ for which $(x,r) \in C$ are called the {\em catalysts of $r$}, and a particular subset of $X$, namely a  set $F$  that has the interpretation as a set of elements that are freely available to the system. Accordingly, $F$ is referred to as a {\em food set}. We write $$r: a_1+ \cdots + a_k[c_1, \cdots, c_r] \rightarrow b_1+\cdots +b_l$$ to denote the reaction that has the reactants $A=\{a_1, \ldots, a_k\}$, the products $B=~\{b_1, \ldots, b_l\}$, and the catalysts $\{c_1, \ldots, c_r\}$.

Let $\rho(r)$ denote the set of reactants of $r$ (i.e., $A$, ignoring multiplicities), and let $\pi(r)$ denote the products of $r$ (i.e.,  $B$, ignoring multiplicities)\footnote{It is assumed that $\rho(r), \pi(r) \neq \emptyset$ for all $r \in R$.}.   Moreover, for a subset $R'$ of $R$, it is convenient to let  $\pi(R') = \bigcup_{r\in R'} \pi(r)$ denote the set of products of the reactions in $R'$.  
 
A subset $R'$ is {\em  F-generated} if the reactions in $R'$ can be placed in some linear order $r_1, r_2, \ldots, r_k$ so that  $\rho(r_1) \subseteq F$ and for all $j$ between 2 and $k$ we have $\rho(r_j) \subseteq F \cup \pi(\{r_1, \ldots, r_{j-1}\})$.  In other words, the reactions in $\R'$ are F-generated if they can proceed in some order so that the reactant(s) of each reaction are available by the time they are first required. We call such an ordered sequence of $R'$ an {\em admissible ordering}. 

Finally, given a CRS $\Q = (X,R,C,F)$, we say that a subset $R'$ of $R$ is a {\em  RAF (Reflexively Autocatalytic and F-generated set)}   if $R'$ is nonempty and is  $F$-generated and, in addition, each reaction $r \in R'$ is catalysed by at least one element in $F \cup \pi(R')$. 
For any CRS $\Q$, let  $\C^{\rm RAF}_\Q$ denote the set of RAFs for $\Q$.

\bigskip

\begin{example}
\label{sample}
Consider the CRS $\Q =(X,R, F, C)$ for which  $X=\{f, f', x,y,z\}$, $F = \{f, f'\}$ and the set $R$ of reactions (with a catalyst indicated in square brackets) is given by:
$$r_1: f[f'] \rightarrow x; \mbox{ } r_2: x [y] \rightarrow z; \mbox{ and } r_3: x+f [z] \rightarrow y.$$
In this case,  $R$ has exactly  two  admissible orderings ($r_1, r_2, r_3$ and $r_1, r_3, r_2$),  and $\C^{\rm RAF}_\Q = \{\{r_1\}, \{r_1, r_2, r_3\}\}$.   
\end{example}

\bigskip


\subsection{The maxRAF interior operator}

A basic result  is that when a  CRS $\Q$ has a RAF,  it has has a unique maximal RAF (which is the union of all the RAFs for $\Q$),  denoted ${\rm maxRAF}(\Q)$ \cite{hor}. 
For any subset $R'$ of $R$, let $\Q|R'$ be the CRS $(X, R', C', F)$, where $C'$ is the restriction of $C$ to $X \times R'$, and 
let $\varphi_\Q:2^R \rightarrow 2^R$ be the following function:

\begin{equation}
\varphi_\Q(R') = 
\begin{cases}
{\rm maxRAF}(\Q|R'), & \mbox{if $\Q|R'$ has a RAF};\\
\emptyset, & \mbox{otherwise}. 
\end{cases}
\label{varphieq}
\end{equation}

To describe how $\varphi_\Q$ can be viewed as an interior operator, we will first recall some further terminology.
Given a subset $R'$ of reactions $R$, a subset $W$ of $X$ is said to be $R'$-closed  if the following property holds:
\begin{itemize}
\item
If a reaction $r$ in $R'$ has all its reactants in $W$ (i.e. $\rho(r) \subseteq W$), then all the products of $r$ are also in $W$ (i.e., $\pi(r) \subseteq W$). 
\end{itemize}
The union of two $R'$-closed sets need not be $R'$-closed; nevertheless, given a nonempty subset $W_0$ of $X$, there is a unique minimal $R'$-closed set containing $W_0$, denoted ${\rm cl}_{R'}(W_0)$. This can be computed in polynomial time in the size of the system by  constructing a nested increasing sequence of subsets of elements $$W_0 \subset W_1, \ldots \subset W_k =W_{k+1} \subseteq X$$ where:
$$W_{i+1} = \blue{W_i \cup} \{x \in X: \exists r\in R': \rho(r) \subseteq W_{i}, x \in \pi(r)\}, \mbox{ for $i\geq 0$}.$$ We then have ${\rm cl}_{R'}(W_0) =  W_k$ (note that $k$ is the first value of $i$ for which $W_i=W_{i+1}$).  If we now take $W_0=F$, it turns out that any subset $R'$ of $R$ is $F$-generated if and only if $\rho(r) \subseteq {\rm cl}_{R'}(F)$ for all $r \in R'$  \cite{ste2}; moreover, $R'$ is a RAF if  $R' \neq \emptyset$ and for each $r \in R'$, the reactants of $r$ and at least one catalyst of $r$ is present in ${\rm cl}_{R'}(F)$. 
This allows us to express $\varphi_\Q$ as an operator of the form $\psi_\lambda$, where $\lambda$ is a function on $2^R$ that satisfies the interior operator properties $(I_1)$ and $(I_2)$.

Let $\lambda_\Q: 2^R \rightarrow 2^R$ be the function defined by:
 $$\lambda_\Q(R') = \{r \in R': \rho(r) \subseteq {\rm cl}_{R'}(F) \mbox{ and } \exists x \in {\rm cl}_{R'}(F): (x,r) \in C\}.$$
The  function $\lambda_\Q$ clearly satisfies conditions ($I_1$) and ($I_2$). If we  
recall the definition of $\psi_\lambda$ from Eqn.~(\ref{lambda}), the maxRAF operator has a representation in the following result from  \cite{ste}. 

\bigskip

\begin{proposition}
For any CRS $\Q=(X,R,C,F)$, the map $\varphi_\Q: 2^R \rightarrow 2^R$ is precisely  the interior operator $\psi_\lambda$ for $\lambda = \lambda_\Q$.
\end{proposition}

\bigskip

This identity  ($\varphi_\Q = \psi_\lambda$) allows for a polynomial-time algorithm to compute $\varphi_\Q$  ({\em c.f.} \cite{ste} and the references therein). In particular, a nonempty subset $R'$ of $R$ is a RAF if and only if $\varphi_\Q(R') = R'$.  Some new and interesting  algebraic (semigroup) properties of the map $\varphi_\Q$ were established recently in \cite{lou2} (see also \cite{lou1}, which considers a more general notion than a RAF, corresponding to `pseudo-RAFs' in the RAF literature, and which we do not explore further in this paper).

\subsection{RAFs in elementary CRS systems}
\label{ele}

At this point, it is helpful to consider a very special type of catalytic reaction system. A CRS $\Q=(X,R,C,F)$ is said to be {\em elementary} if each of its reactions has all its reactants in the present food set (formally, $\rho(r) \subseteq F$ for each $r \in R$). 

Given an elementary CRS $\Q=(X,R,C,F)$, define a digraph $D(\Q) =(V, A_\Q) $ to have vertex set $R$ and an arc from $r$ to $r'$ ($r \neq r'$) if a product of $r$ catalyses $r'$;  in addition, we place an arc from $r$ to itself if either a product of $r$ or an element of $F$ catalyses $r$.

The following result  is easily verified from the definitions (or see \cite{ste2}, Theorem 2.1) and describes the set of RAFs of an elementary CRS $\Q$  (i.e., $\C^{\rm RAF}_\Q$) in terms of the fixed sets of the interior operators arising from digraphs (from Section~\ref{digraphs}, and recalling the definition of $\C(D)$).    This will be applied in the next section. 

\bigskip

\begin{lemma}
\label{lemon}
$\C^{\rm RAF}_\Q \blue{\cup \{\emptyset\}}= \C(D(\Q))$.
\end{lemma}

An immediate consequence of this lemma and Proposition~\ref{propro}(ii) is the following.

\bigskip

\begin{corollary} 
If $\Q$ is an elementary CRS which has a RAF, then for any two RAFs of $\Q$ (say, $R', R''$) if  $R' \subsetneq R''$, then either $R''\setminus R'$ is a RAF for $\Q$ or there is some reaction  $r \in R'' \setminus R'$ for which $R''\cup \{r\}$ is a RAF for $\Q$.
\end{corollary}

\bigskip

Note  that this  corollary can fail without the assumption that $\Q$ is elementary;  Example~\ref{sample} provides a counterexample for the two RAFs $R'=\{r_1\}$ and $R'' = R= \{r_1, r_2, r_3\}$.  If  one removes the `elementary' restriction on a CRS, the class of possible set systems that can be realised as RAFs of some suitably chosen CRS becomes larger and less tractable. We investigate this further in the next section, where we will apply Lemma~\ref{lemon} and the earlier Proposition~\ref{propro}(iii).

\subsection{Representing an interior operator as a RAF operator}
\label{final}

The main results in RAF theory that are generic (i.e., which hold regardless of the particular choices or restrictions on $F$, $X$, $R$ or $C$ in $\Q$) can be established by using only the property that the maxRAF operator $\varphi_\Q$ is a (efficiently computable) interior operator  (see \cite{ste}). This raises the question as to whether theorems that hold true for all RAFs can always be established from (just) this generic property.  In other words, can every interior operator on every finite set $Y$ be realised as the maxRAF operator associated with a  suitably chosen catalytic reaction system $\Q= (X, R, C, F)$ in which $Y$ is identified (via a bijection) with the set $R$ of reactions in $\Q$. We  show that the answer is `no' by describing a generic result in RAF theory that  is not a consequence of the interior operator property of the maxRAF operator.

More precisely, we say that $\psi$ has a {\em RAF-realisation}  if there exists a CRS $\Q=(X,R, C, F)$ and a bijection $b: Y \rightarrow R$ such that for each $Y' \in 2^Y$ we have:
$$\psi(Y') = \beta^{-1} \circ \varphi_\Q(R')$$ where $R' =\beta(Y')$ and where $\beta:2^Y \rightarrow 2^R$ is the natural bijection induced by $b$.
 In other words, the diagram shown commutes for each $Y'\subseteq Y$.  Note that no restriction is placed on the sets $X, F$, and $C$ in $\Q$; in particular, they  could be arbitrarily large sets. 
 \begin{figure}[htb]
\centering
\includegraphics[scale=1.0]{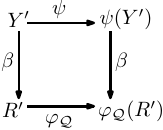}
\label{figcom}
\end{figure}

We now show that such a realisation is not always possible, as described in Proposition~\ref{negthm}(ii) below. For this result, an  {\em irreducible RAF} (iRAF)  for a CRS $\Q= (X, R, C, F)$ is a RAF $R'$ with the property that it contains no (nonempty) RAF as a proper subset (i.e., $\varphi_\Q(R')=R'$ and $\varphi_\Q(R'\setminus \{r\}) = \emptyset$ for all $r \in R'$). 

\bigskip

\begin{proposition} 
\label{negthm}
\mbox{}
\begin{itemize}
\item[(i)] For any integer $k\geq 3$ and any CRS $\Q= (X, R, C, F)$ with $|R|\geq k^3-3k^2+4k$, not all subsets of $R$ of size $k$ are iRAFs.
\item[(ii)]
For any finite set  $R$ of size at least  $12$, there exists an interior operator $\psi$ on $2^R$ that does not have a RAF-realisation.
\end{itemize}
\end{proposition}

{\em Proof of Proposition~\ref{negthm}:}
{\em Part (i): } Let $m=(k^2-3k+3)$, and suppose that $|R| \geq km$ and every subset of $R$ of size $k$ is an iRAF; we will derive a contradiction. 
Since $|R| \geq km$ there exist $m$  disjoint subsets of $R$ of size $k$, call them $R_1, \ldots, R_m$.
Since these are subsets of $R$ of size $k$ they are iRAFs for $\Q$.
Now, any RAF requires at least one reaction to have all its reactants in the food set $F$ (this can easily been verified by considering the first reaction in any admissible ordering of the reactions in a RAF). Thus,  we can select one such reaction $r_i$ from $R_i$ (for each $i$), to obtain a set $R_k=\{r_1, \ldots, r_m\}$ of $m$ (distinct) reactions, with each reaction in  $R_m$ having all its reactants in $F$.
Consider the CRS $\Q_m = (X, R_m, C_m, F)$ by restricting $R$ to $R_m$ and restricting $C$ to $C_m=\{(x,r) \in C: r \in R_m\}$. This is an elementary CRS,  and so, by Lemma~\ref{lemon}, the set of RAFs of $\Q_m$ is equal to \blue{$\C(D(\Q_m)) \setminus \{\emptyset\}$ (where $D(\Q))$ is defined as Section~\ref{ele})}.   Since each subset of $R$ of size $k$  is an iRAF of $\Q$ (and noting that $m \geq k$), it follows that 
\blue{$\C(D(\Q_m))\setminus \{\emptyset\}$} contains all subsets of $R_m$ of size $k$, and no subsets of size less than $k$, and so we can apply Proposition~\ref{propro}(iii)    (with $Y=R_m$) to deduce that $m=|R_m| \leq 1+(k-1)(k-2)$.  But this contradicts the inequality $m  =|R|/k= k^2-3k+4 >  1+(k-1)(k-2)$.

\bigskip

{\em Part (ii):} Put $k=3$ in Part (i) and consider the following map $\psi: 2^R \rightarrow 2^R$:
$$\psi(R'), = \begin{cases}
R', & \mbox{ if }|R'|\geq 3; \\
\emptyset, & \mbox{ if } |R'| \leq 2.
\end{cases}
$$
It is easily verified that $\psi$ satisfies properties ($I_1$), ($I_2$) and ($I_3$) and so is  an interior operator, but  $\psi$ has no RAF-realisation by Part (i).
\hfill$\Box$

\bigskip

\noindent{\bf Remarks:}
\begin{itemize} 
\item The condition that $k\geq 3$ is required in Proposition~\ref{negthm}(i) since for $k\leq 2$ it is easy to construct CRS systems with an arbitrarily large set of reactions and with all subsets of $R$ of size $k$ being iRAFs (based on the second remark following Proposition~\ref{propro}).

\item Note also that the value 12 in Proposition~\ref{negthm}(i) (when  $k=3$) can be reduced to 4 if one restricts to RAF representations within {\em elementary} CRS systems. However,  without that restriction, Proposition~\ref{negthm}(i) does not hold if 12 is replaced by 4. An example is provided by the CRS $\Q$ consisting of $X=\{f, c_1,c_2, c_3, \gamma, x,y,z\}, F=\{f\}$ and $R$ comprising the four catalysed reactions:

$$r_1 : f [c_3,\gamma]\rightarrow  x+y+c_1$$
$$r_2 : f [c_1,\gamma]\rightarrow y+z+c_2$$ 
$$r_3 : f [c_2,\gamma]\rightarrow x+z+c_3$$
$$ r_4 : x+y+z  [w]\rightarrow w+\gamma$$
For this system, each of the four subsets of $R$ of size 3 is an iRAF of $\Q=(X,R,C,F)$.
\end{itemize}

\blue{It is possible that the value of 12 in Proposition~\ref{negthm}(i) (when  $k=3$) could be reduced further (or that the lower bound of 4 provided by the example above could be increased), however this would require more elaborate arguments.}

\section{Concluding comments}
Proposition~\ref{propro} provides set-theoretic necessary conditions for a union-closed collection of sets to be realisable by a digraph. A natural question is whether there is a set-theoretic characterisation of the class of union-closed sets to be realisable by a digraph.  A more difficult task would be to  characterise the set systems that are realisable as the RAFs of some CRS.  Related to the (still open) union-closed conjecture \cite{bal}, is the question of whether there is always a reaction that lies in at least half the RAFs (for either an elementary or general CRS).   Although we have focused on applications of interior operators arising from digraphs to autocatalytic networks, other  properties of interior operators realisable by graph-based processes may also be relevant to various applications (e.g. in investigating the fixed sets present within digraph models of neuronal networks of the type discussed in \cite{gri})).

\section{Acknowledgements} I thank Dimitri Loutchko for some helpful comments on an earlier version of this manuscript, \blue{and the two anonymous reviewers further comments.}

\section{Declarations}

{\bf Competing interests} The author declares no competing interests and that this work was not supported by any external grant funding. 

\section{Appendix}

{\em Proof of Lemma~\ref{helps}}

{\em Part (i):} The first claim  follows immediately from Condition ($I_3$).
For the second claim, observe that the term on the right is a subset of $\psi(X)$ since every set $U$ in $F_\psi(X)$  is a subset of $X$ and so we can apply Condition ($I_2$). However $\psi(X)$ is also in $F_\psi(X)$ (by Condition ($I_3$)), so the $\psi(X)$ is a subset of the right-hand side. 

{\em Part (ii):} If $A \in W$, then $A=\psi(A)\subseteq \psi(\bigcup W)$ by $(I_2)$,  so $\bigcup W \subseteq \psi(\bigcup W)$.
Since $\psi(\bigcup W) \subseteq \bigcup W$ (by $I_1$) it follows that $\psi(\bigcup W) = \bigcup W$ and so $\bigcup W \in F_\psi(X)$, as claimed.

{\em Part (iii): } If $\C$ is a collection of fixed sets of some interior operator, then $\emptyset \in \C$, and $\C$ is union-closed by Part (ii). 
Suppose that $\C$ has these properties, and consider $\psi_{\C}$. This function clearly satisfies ($I_1$) and ($I_2$). To verify Condition ($I_3$), observe that 
the union closure condition on $\C$ implies that $\bigcup\{U \in \C: U \subseteq X\}$ is a set $U'$ in $\C$, and so 
$$\psi_{\C}(\psi_{\C}(X)) = \psi_{\C}(U') = \bigcup\{U'' \in \C: U'' \subseteq U'\}= U' = \psi_{\C}(X),$$
as required.  Moreover, the fixed set of $\psi_{\C}$ is $\C$, since if $X \in \C$, then $\psi_{\C}(X) = \bigcup\{U \in \C: U \subseteq X\} = X$, and if  $\psi_{\C}(X) =X$ then since $\C$ is union-closed, this implies that $X \in \C$. 
For the uniqueness claim, suppose that $\psi$ has $F_\psi = \C$. From Part (i), $\psi(X)= \bigcup_{U \in F_\psi(X)} U$ and 
$F_\psi(X) = \{U \in \C: U \subseteq X\}$ thus $\psi=\psi_\C.$
\hfill$\Box$

\bigskip

\bigskip

\noindent{\em Proof of Proposition~\ref{propro}}

{\em Part (i):} Suppose that $X, X' \in \C(D)$, and $v \in X \cup X'$. Without loss of generality, we may suppose that $v \in X$. Then $v$ has strictly positive in-degree in $D|X$, and any arc of the form $(x, v)$ with $x \in X$ is also present in $D|(X \cup X')$. Thus $X \cup X' \in \C(D)$. 
For the second claim, if $v_1, \ldots, v_r = v_1$ is a directed cycle in $D$, then the set of vertices in this cycle lies in $\C(D)$.  Conversely, if $D$ is acyclic, then so too is $D|X$ for any subset $X$ of $Y$, and since every finite acyclic directed graph has a vertex of in-degree 0, it follows that $X\not\in \C(D)$. 

\bigskip

{\em Part (ii):}
Suppose there is no vertex $w \in W\setminus U$ for which $U \cup \{w\} \in \C(D)$. Then there is no arc from any vertex in $U$ to a vertex in $W\setminus U$. However, every vertex in $W\setminus U$ has an incoming arc from some vertex in $W$, and therefore, it has an incoming arc from some vertex in $W\setminus U$. Thus $D|(W\setminus U)$ has the property that every vertex in this induced graph has in-degree at least 1, so $W\setminus U \in \C(D)$.

\bigskip

{\em Part (iii):}
Let $D=(V, A)$, and  suppose that $\C_k(D) = \binom{Y}{k}$ and $\C_{j}(D) = \emptyset$ for  all $1\leq j<k$, where $k\geq 3$. We first show that this implies that $d^+_{D}(v) \leq k-2$
for each  vertex $v \in Y$.  To see this, suppose that $d^+_{D}(v') \geq k-1$ for some element $v' \in Y$; we will derive a contradiction.  Observe that  $(v',v') \notin A$ (otherwise $\{v'\} \in \C_1(D) = \emptyset$) and so there is a subset $X'$ of  size at least $k-1$ for which $(x',v') \in A$ for each $x' \in X'$. Let $X''$ be any subset of $X'$ of size exactly $k-1$. 
Since  $\C_{k-1}(D)=\emptyset$ and $|X''|=k-1$, at least one element $x'' \in X''$ has no incoming arc from any other vertex in $X''$, which means that $(v',x'') \in A$, since  $X'' \cup \{v'\} \in \C_k(D)$. On the other hand, $(x'', v') \in A$ (by definition of $X''$), which implies that  $\{v', x''\} \in \C_2(D)$, providing the required contradiction since $\C_2(D)=\emptyset$ (since $k\geq 3$). 
Thus each vertex $v$ in $D$ has in-degree at most $k-2$, as claimed. 

If we now let $n=|Y|$ then,  since $|A| = \sum_{v \in Y}d_D^+(v)$, we obtain:
\begin{equation}
\label{likely}
|A| \leq (k-2)\cdot n.
\end{equation}
Now consider the set $\Omega$ of pairs $(S, a)$ where $S$ is a subset of $k$ vertices from $Y$, and $a$ is an arc between any two vertices of $S$. Formally, 
$$\Omega = \{(S, a): S \in \binom{Y}{k}; a \in A \cap (S \times S)\}.$$
We count this set in two ways. Since $n= |Y|$, the number of choices for $S$ is $\binom{n}{k}$. Moreover, for each such set $S$, there are precisely $k$ arcs that form a cycle involving $k$ elements in $S$, since: (a) if any more arcs were present between the vertices of $S$ then  a set in $\C_{j}(D)$ for some $j<k$ would appear, and (b) if no cycle was present involving all elements of $k$, then $S$ would not lie in $\C_k(D)$, and both of these two possibilities are excluded by the two assumptions stated in Part (iii). 
In summary,
\begin{equation}
\label{likely2}
|\Omega| = \binom{n}{k} \cdot k
\end{equation}
We can  also count $\Omega$ by first selecting an arc $(u,v)$ from $A$ and counting the number of sets $S \in \binom{Y}{k}$ that contain $u$ and $v$. By the assumptions in Part (iii), each subset of $Y$ of size $k$ induces a unique cycle through all the vertices (and with no other arcs present between the vertices), so the number of sets $S$ that can be chosen for $(u,v)$ is $\binom{n-2}{k-2}$. Thus we have: 
\begin{equation}
\label{likely3}
|\Omega| =\binom{n-2}{k-2} \cdot |A|
\end{equation}
Combining Eqns. (\ref{likely}), (\ref{likely2}) and (\ref{likely3}) gives:
$$\binom{n}{k}k \leq \binom{n-2}{k-2}(k-2) n,$$
which simplifies to $n \leq 1+(k-1)(k-2)$, as claimed.  
\hfill$\Box$

\bigskip

\noindent {\em Proof of \blue{Proposition}~\ref{lem2}:}

Let $\A=\{U \subset Y: |U|=\lfloor n/2\rfloor\}$, which is an antichain in the poset $2^Y$ (partially ordered by set inclusion) of size $\binom{n}{\lfloor n/2\rfloor}$ ($\A$ is also a largest antichain by Sperner's theorem).
Let  $S$ be a subset of $\A$,
and let $\C[S]$ be the collection of subsets of $Y$ consisting of $\emptyset$, \blue{the sets in} $S$, and \blue{all possible unions of} the sets from $S$.  In this case, $\C[S]$ satisfies the conditions of Lemma~\ref{helps} (iii) and so there is a unique interior operator $\psi_{\C[S]}$ that has the fixed set $\C[S]$.  Moreover, the collection of minimal \blue{nonempty} fixed sets of $\psi_{C[S]}$ is precisely the sets in $S$,  so if $S \neq S'$, then $\psi_{\C[S]} \neq \psi_{\C[S']}$. Since there are \blue{$2^{\binom{n}{\lfloor n/2\rfloor}}$} choices for $S$, this completes the proof.
\hfill$\Box$

\bigskip

\end{document}